\documentstyle[aps,prb,multicol,ifthen,epsfig]{revtex}

\newcommand{\wide}[2]{                                                                                             %
\end{multicols}                                                                                                                 %
\widetext                                                                                                                            %
\noindent                                                                                                                           %
\ifthenelse{\equal{#1}{t}}                                                                                              %
{}                                                                                                                                           %
{                                                                                                                                            %
\raisebox{0.1in}[0in][0.02in]{$\rule{3.575in}{0.002in}                                            %
\rule{0.002in}{0.08in}$}                                                                                                  %
}                                                                                                                                            %
#2                                                                                                                                         %
\ifthenelse{\equal{#1}{b}}                                                                                             %
{}                                                                                                                                           %
{                                                                                                                                            %
{\raisebox{-0.1in}[0in][0.02in]                                                                                       %
{\hspace{3.575in}$\rule{0.002in}{0.08in}                                                                   %
\rule[0.08in]{3.575in}{0.002in}$}                                                                                   %
}                                                                                                                                             %
}                                                                                                                                             %
\begin{multicols}{2}                                                                                                         %
\noindent                                                                                                                            %
}                                                                                                                                             %

\begin{document}

\title{Localization corrections to the anomalous Hall effect 
in a ferromagnet}
\author{V.~K.~Dugaev$^{1,2}$\cite{email}, A.~Cr\'epieux$^1$, and P.~Bruno$^1$}
\address{$^1$Max-Planck-Institut f\"ur Mikrostrukturphysik,
Weinberg 2, D-06120 Halle, Germany\\
$^2$Institute for Materials Science Problems, Ukrainian Academy of Sciences,\\ 
Vilde 5, 58001 Chernovtsy, Ukraine}
\date{Received \today }
\maketitle

\begin{abstract}
We calculate the localization corrections to the anomalous 
Hall conductivity related to the contribution of spin-orbit scattering 
into the current vertex (side-jump mechanism). We show that
in contrast to the ordinary Hall effect, there exists a nonvanishing localization
correction to the anomalous Hall resistivity. The correction to
the anomalous Hall conductivity vanishes in the case of side-jump
mechanism, but is nonzero for the skew scattering.
The total correction to the nondiagonal conductivity related to both
mechanisms, does not compensate the correction to the diagonal conductivity. 
\end{abstract}

\begin{multicols}{2}

The Anomalous Hall (AH) effect can be observed in magnetically ordered
metals or semiconductors without external magnetic field.\cite{hurd,chien}
The key point of any explanation of this effect is the presence of spin-orbit
(SO) interaction, which breaks the symmetry to spin rotations.

The theory of AH effect has been developed in numerous 
works\cite{luttinger,nozieres,abakumov,fert,kondor,vedyaev84}.
More recently, the interest to this effect is 
growing\cite{vedyaev,zhang,hirsch,bulgakov,ye,crepieux} due to the 
importance of the spin polarization and spin-orbit interaction for transport 
properties of materials and structures of spin 
electronics.\cite{prinz,fiederling,awschalom,ohno} 
Besides, the measurement of AH effect is proved to be a useful 
tool to determine the magnitude 
of magnetization in structures with magnetic layers.\cite{ohno}
 
Usually, two relevant mechanisms are distinguished - a skew 
scattering\cite{smit,nozieres} and a side-jump effect.\cite{berger,lyo}
It is commonly believed that
the first mechanism prevails in low-resistivity metals, whereas the other one 
(side-jump) can be more significant for metal alloys or
semiconductors with much larger
resistivity.\cite{hurd,chien}  

The theory of localization corrections to the conductivity and Hall conductivity
is developed in details for nonmagnetic metals and heavily doped 
semiconductors\cite{alt82,alt85,lee85,bergmann}, but not for magnetically 
ordered materials. In our recent works\cite{dugaev} we analysed some effects
related with localization and interaction corrections in ferromagnets and 
in multilayer structures with thin magnetic layers.
 
The role of quantum corrections (both localization and exchange-interaction) 
to AH effect 
has been considered theoretically in Ref.~[\onlinecite{langenfeld}] but only in the case 
of skew scattering. In the present work we consider the localization corrections
in the framework of side-jump mechanism. Also, we revisit the calculation of localization
corrections for the skew scattering in the model of itinerant magnetism and confirm
the result\cite{langenfeld} found in a model of impurities with ordered magnetic
moments. 
We show that the results for the side-jump and skew scattering are quite different.

We consider a ferromagnet with a strong exchange 
magnetization ${\bf M}$ oriented along the axis $z$, 
and a SO relativistic term (we put $\hbar =1$)
$$
H=\int d^{3}r\, 
\psi ^\dag ({\bf r})\left[ -\frac{\nabla ^{2}}{2m^*}-M\sigma _{z}
\right.
$$
$$
\left.
-\, \frac{i\, \lambda _0^2}{4} 
\left( {\bf \sigma }\times \nabla V({\bf r})\right) \cdot \nabla 
+V({\bf r})\right] \psi ({\bf r})\; ,
\eqno (1)
$$
where $m^*$ is the electron effective mass,
$\lambda _0$ is a constant, which measures the strength of the SO
interaction, $V({\bf r})$ is a random potential  created by impurities or defects, 
$\sigma =(\sigma _x, \sigma _y, \sigma _z)$ are the Pauli matrices, 
and $\psi ^\dag \equiv \left( \psi ^\dag _\uparrow ,\, 
\psi ^\dag _\downarrow \right) $ is the spinor field, corresponding to
electrons with spin up and down orientations. The constant $\lambda _0$ has
the dimensionality of length. For non-relativistic electrons in vacuum,
$\lambda _0$ is equal to $\lambda _c/2\pi $, where $\lambda _c=2\pi /m_0c$  
is the Compton wavelength of electron and $m_0$  is the 
free electron mass. 

We assume that the potential $V({\bf r})$ is short-ranged, with zero mean value,
$\langle V({\bf r})\rangle =0$, where the angle brackets mean the  configurational
averaging over all realizations of $V({\bf r})$.  
We shall characterize this potential by its second, $\gamma _2$, and 
third, $\gamma _3$, momenta, denoting
$\langle V({\bf r}_1)\, V({\bf r}_2)\rangle =\gamma _2\, \delta({\bf r}_1-{\bf r}_2)$ and  
$\langle V({\bf r}_1)\, V({\bf r}_2)\, V({\bf r}_3)\rangle 
=\gamma _3\, \delta ({\bf r}_1-{\bf r}_3)\, \delta ({\bf r}_2-{\bf r}_3)$.

It should be emphasized that the constants $\gamma _2$ and $\gamma _3$ are  
parameters, characterizing not only the strength of the disorder potential, but
also the statistical properties of the random field. 
When the potential $V({\bf r})$ is created by impurities, distributed
randomly at some points ${\bf R}_i$, we have $V({\bf r})=\sum _iv({\bf r}-{\bf R}_i)$.
It results in $\gamma _2=N_i\, v_0^2$ and $\gamma _3=N_i\, v_0^3$, where $N_i$ is the
impurity concentration, and $v_0$ is the matrix element of the short-ranged
potential of one isolated impurity, $v({\bf r}-{\bf R}_i)=v_0\, \delta ({\bf r}-{\bf R}_i)$. 
In the case of purely Gaussian potential, we should take $\gamma _3=0$. 

\begin{figure}
\hskip2cm
\epsfig{file=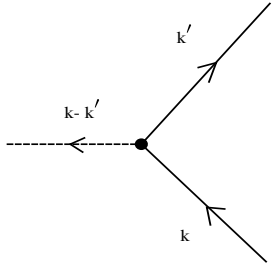,height=4cm,width=4cm}
\caption{Anomalous vertex for the coupling to electromagnetic field with SO 
interaction. The dashed line is for the impurity scattering and the filled black
circle is for the external electromagnetic field.}
\end{figure}

Calculating the matrix elements of the Hamiltonian (1) in momentum 
representation, we obtain
$$
H=\sum _{\bf k}\, \psi ^\dag _{\bf k}
\left( \frac{k^{2}}{2m^*}-M\sigma _{z}\right) \psi _{\bf k}
$$
$$
+\, \sum _{{\bf kk}'}\psi ^\dag _{\bf k}\, V_{{\bf k}-{\bf k}'}
\left[ 1+\frac{i\, \lambda _0^2}{4}\left( {\bf k}
\times {\bf k}^\prime \right) \cdot {\bf \sigma }\right] \psi _{{\bf k}^\prime }\; ,
\eqno (2)
$$
where $V_{\bf k}$ is the Fourier transform of the potential $V({\bf r})$.
The second term in Eq.~(2) describes the SO scattering from impurities.

To find the expression for current density operator ${\bf j}(t)$, we switch 
on an electromagnetic field ${\bf A}(t)$ in a gauge-invariant way, 
${\bf k}\rightarrow ({\bf k}-e{\bf A}/c)$, and calculate the derivative
$$
j_\alpha =-c\, \frac{\delta H}{\delta A_\alpha }\; ,
\eqno (3)
$$
which gives us

$$
j_\alpha =\sum _{{\bf kk}'}\psi _{\bf k}^\dag
\left[ \frac{e}{m^*}\left( k_\alpha -\frac{eA_\alpha }{c}\right) \delta _{{\bf kk}'}
\right. 
$$
$$
\left.
+\, \frac{ie\, \lambda _0^2}{4}\, V_{{\bf k}-{\bf k}'}\, \epsilon _{\alpha\beta\gamma}\,
(k_\beta ^\prime -k_\beta )\, 
\sigma _\gamma \right] \psi _{{\bf k}'}\; ,
\eqno (4)
$$ 
where $\epsilon _{\alpha\beta\gamma}$ is the unit antisymmetric tensor.

According to Eq.~(4), the SO interaction contributes to the current 
vertex in the Feynman diagrams of the conductivity tensor.\cite{agd} 
The additional anomalous vertex (second term in Eq.~(4)) can be 
presented by a three-leg vertex, Fig.~1, where the dashed line corresponds
to the interaction $V_{{\bf k}-{\bf k}'}$ with impurities, and the black point implies 
the coupling to the external electromagnetic field.

Calculating the Feynman diagrams for the off-diagonal (Hall) conductivity, 
Fig.~2, we find (an additional factor '2' comes from the contributions of right  
vertices)
$$
\sigma _{xy}^{(sj)}
=-\frac{ie^2\lambda _0^2\, \gamma _2}{4\pi m^*}\, {\rm Tr}
\sum _{{\bf kk}'}\sigma _z\, k_y^2\, G_{\bf k}^R\, G_{\bf k}^A\,
\left(G_{{\bf k}'}^R-G_{{\bf k}'}^A\right) \; ,
\eqno (5)
$$
\begin{figure}
\hskip2cm
\epsfig{file=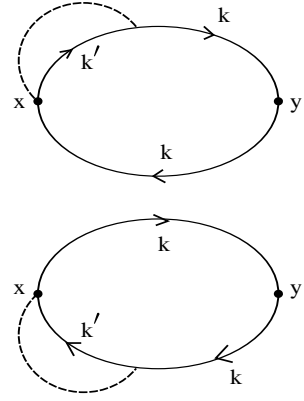,height=6cm,width=4.5cm}
\caption{Feynman diagrams for AH conductivity $\sigma _{xy}^{(sj)}$ 
(side-jump mechanism).}
\end{figure}
\noindent
where the retarded (R) and advanced (A) Green functions at the Fermi
surface are diagonal matrices
$$
G_{\bf k}^{R,A}={\rm diag}\left(
\frac1{\mu -\varepsilon _{\uparrow }(k)\pm i/2\tau _\uparrow }\; ,\;
\frac1{\mu -\varepsilon _{\downarrow }(k)\pm i/2\tau _\downarrow }
\right) .
\eqno (6)
$$
Here $\varepsilon _{\uparrow ,\downarrow }(k)=k^2/(2m^*)\mp M$ are the energy
spectra of spin-up and spin-down electrons, respectively, $\mu $ is the chemical
potential, and
$\tau _{\uparrow ,\downarrow }$ are the corresponding relaxation
times. 
The relaxation times are
determined by the scattering from the random potential, and they are equal to
$\tau _{\uparrow ,\downarrow }
=\left( 2\pi \nu _{\uparrow ,\downarrow }\gamma _2\right) ^{-1}$,
where $\nu_\uparrow $ and $\nu _\downarrow $ are the densities of states
for spin-up (majority) and spin-down (minority) electrons at the Fermi level.

\begin{figure}
\psfig{file=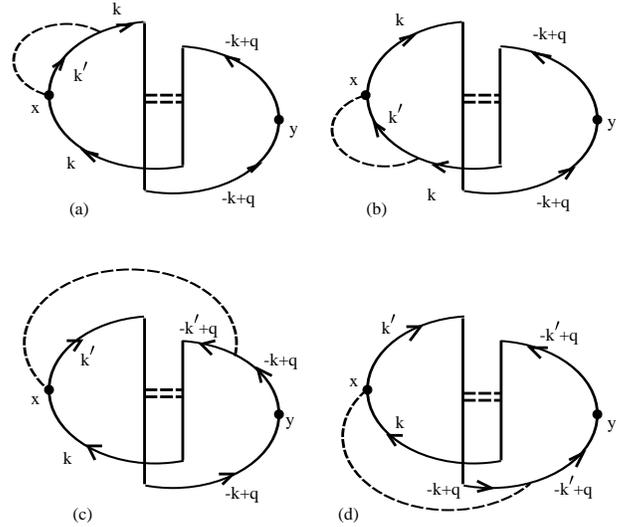,height=7.5cm,width=8.5cm}
\caption{Localization corrections to $\sigma _{xy}^{(sj)}$ due to the Cooperons.}
\end{figure}

\begin{figure}
\hskip1.5cm
\epsfig{file=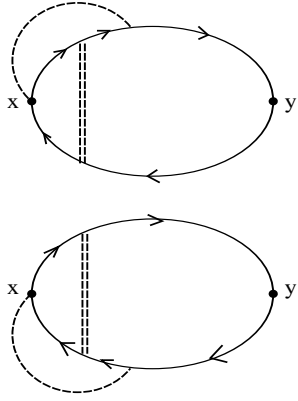,height=6cm,width=5cm}
\caption{Localization corrections to $\sigma _{xy}^{(sj)}$ due to the Diffusons.}
\end{figure}

After calculating the integrals in Eq.~(5), we find the side-jump AH
conductivity\cite{crepieux} (in final formulae we restore $\hbar $ and use
the electron parameters at the Fermi surfaces)
$$
\sigma _{xy}^{(sj)}=\frac{e^2}{6\hbar }\, \lambda _0^2\, 
\left( \nu_\downarrow \, \hbar k_{F\downarrow }\, v_{F\downarrow }
-\nu _\uparrow \, \hbar k_{F\uparrow }\, v_{F\uparrow }\right) \; ,
\eqno (7)
$$
where $k_{F\uparrow ,\downarrow }$ and $v_{F\uparrow ,\downarrow }$ 
are the momenta and velocities of majority and minority electrons at the Fermi
surfaces, respectively.

Now we consider the localization corrections to $\sigma _{xy}^{(sj)}$. 
They can be presented by the loop diagrams with Diffusons and 
Cooperons.\cite{alt82,alt85,lee85}.
Assuming the exchange energy $M$ larger than $1/\tau $, we can restrict ourselves
by considering only triplet Cooperons and Diffusons, with the same orientation
of spins in the particle-particle (Cooperon) or particle-hole (Diffuson) channels.   

There are eight diagrams containing such Cooperons and four diagrams with 
Diffusons, presented in Figs.~3 and 4, respectively (the figures show only diagrams
with left anomalous vertices). 

We calculate first the quantum corrections due to the Cooperons. 
Calculating the first two diagrams of Fig.~3 ($a$ and $b$), we find
$$
\Delta \sigma _{xy}^{(1)}
=\frac{ie^2\lambda _0^2\, \gamma _2}{8\pi m^*}\, 
$$
$$
\times \, {\rm Tr}\sum _{{\bf kk}'{\bf q}}\sigma _zk_y^2
\left( G_{{\bf k}'}^R-G_{{\bf k}'}^A\right)
\left( G_{\bf k}^R\right)^2\left( G_{\bf k}^A\right)^2
C(0,{\bf q})\; ,  
\eqno (8)
$$
where the spin components of the Cooperon\cite{alt82,alt85,lee85}  
are equal to 
$$
C_\sigma (\omega ,{\bf q})=\frac1{2\pi \nu _\sigma \tau _\sigma ^2}\; 
\frac1{-i\omega +D_\sigma q^2+1/\tau _{so,\sigma }+1/\tau _{\varphi ,\sigma }}\; .
\eqno (9)
$$
Here $D_\sigma =v_{F\sigma }^2\tau _\sigma /d\, $ is the diffusion constant of
electrons ($d$ the effective dimensionality\cite{lee85}),
$\tau _{so,\sigma }$ and $\tau  _{\varphi ,\sigma }$ are the spin-orbit 
and phase relaxation times, respectively.  In Eq.~(8) we neglected small momentum
$q\ll k\sim k_F$ in the arguments of Green's functions. 

The calculation of two other diagrams of Fig.~3 ($c$ and $d$) gives us
$$
\Delta \sigma _{xy}^{(2)}=
-\frac{ie^2\lambda _0^2\, \gamma _2}{8\pi m^*}\, 
$$
$$
\times \, {\rm Tr}\sum _{{\bf kk}'{\bf q}}\sigma _z
(k'_y)^2\left( G_{{\bf k}'}^R-G_{{\bf k}'}^A\right)
G_{{\bf k}'}^R\, G_{{\bf k}'}^A\, 
G_{\bf k}^R\, G_{\bf k}^A\,
C(0,{\bf q})\; . 
\eqno (10)
$$

\begin{figure}
\epsfig{file=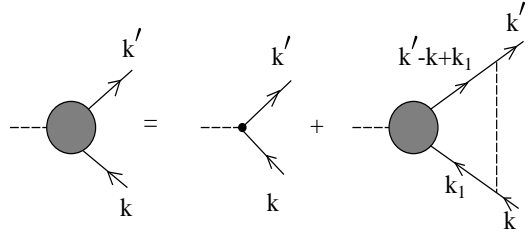,height=3.5cm,width=7.5cm}
\caption{The equation for the renormalized anomalous vertex 
$\Gamma ({\bf k}, {\bf k}^\prime )$.}
\end{figure}

After integrating over ${\bf k}$ and ${\bf k}'$, we find that the contributions 
of all diagrams with Cooperons, Eqs.~(8) and (10), cancel each other exactly.
This result can also be seen by comparing directly Eqs.~(8) and (10) and by
using the property of Green's functions:
$G_{\bf k}^R-G_{\bf k}^A=(-i/\tau )G_{\bf k}^R\, G_{\bf k}^A$. 

The diagrams with Diffusons (Fig.~4) can be taken into account as a renormalization
of the anomalous vertex by impurities. 
As is known, the normal electromagnetic 
vertex without SO correction (first term in Eq.~(4)) can be renormalized only for 
non-pointlike defects\cite{agd}. Here we show that the anomalous vertex is 
renormalized in the case of pointlike defects, too.

The equation for the three-leg vertex has the following form (Fig.~5)
$$
\Gamma ({\bf k},{\bf k}')=\Gamma ^0({\bf k},{\bf k}')
$$
$$
+\, \gamma _2\, \sum _{{\bf k}_1}G_{{\bf k}_1}^A\, 
\Gamma ({\bf k}_1,{\bf k}_1+{\bf k}'-{\bf k})\, G_{{\bf k}'-{\bf k}+{\bf k}_1}^R\; ,
\eqno (11)
$$
where
$$
\Gamma _\alpha ^0({\bf k},{\bf k}')
=\frac{ie\, \lambda _0^2}{4}\, V_{{\bf k}-{\bf k}'}\, \epsilon _{\alpha\beta\gamma}
(k'_\beta -k_\beta )\, \sigma _\gamma \; .
\eqno (12)
$$

In view of Eq.~(12), we can look 
for a solution of Eq.~(11) in the form of matrix in the spin space that depends 
only on the difference of momenta, 
$\Gamma _\alpha ({\bf k},{\bf k}')=\Gamma _\alpha ({\bf k}'-{\bf k})$.
Hence, the solution can be presented as
$$
\Gamma _\alpha ({\bf q})=\Gamma _\alpha ^0({\bf q})
\left[ 1-\gamma _2\, \Pi ({\bf q})\right] ^{-1}\; ,
\eqno (13)
$$
where $\Pi ({\bf q})$ is the diagonal matrix
$$
\Pi ({\bf q})=\sum _{\bf k}G_{{\bf k}+{\bf q}}^R\, G_{\bf k}^A\; .
\eqno (14)
$$

According to (11) and (12), we can present the $x$-component of the vertex 
$\Gamma ({\bf q})$ as
$$
\Gamma _x({\bf q})=q_y\, \Gamma _x^y({\bf q})+q_z\, \Gamma _x^z({\bf q})\, ,
\eqno (15)
$$
where
$$
\Gamma _x^y({\bf q})=\frac{ie\, \lambda _0^2}{4}\, V_{\bf q}\, {\rm diag}\left(
\frac1{D_\uparrow q^2\tau _\uparrow },\,
-\frac1{D_\downarrow q^2\tau _\downarrow }\right) \; ,
\eqno (16)
$$
and only $\Gamma _x^y({\bf q})$ component is needed, since the second term in
Eq.~(15) gives the vanishing contribution to $\Delta \sigma _{xy}^{(sj)}$.

Using (15) and (16), we find the localization corrections to $\sigma _{xy}^{(sj)}$
$$
\Delta \sigma _{xy}^{(sj)}
=\frac{ie^2\lambda _0^2\, \gamma _2}{8\pi m^*}
$$
$$
\times \, {\rm Tr}\sum _{{\bf kq}}\sigma _z
\frac{q_y\, k_y}{D_\sigma q^2\tau _\sigma } 
\left( G_{{\bf k}+{\bf q}}^R+G_{{\bf k}-{\bf q}}^A\right) G_{\bf k}^R\, G_{\bf k}^A\; .
\eqno (17)
$$
Taking into account that very small momenta $q$ can be essential for the Diffuson, 
$q<(D\tau )^{-1/2}\ll k\sim k_F$, we can present (17) in the form
$$
\Delta \sigma _{xy}^{(sj)}
=-\frac{ie^2\lambda _0^2\, \gamma _2}{8\pi m^*}
$$
$$
\times \, {\rm Tr}
\sum _{{\bf kq}}\sigma _z
\frac{q_y^2}{D_\sigma q^2\tau _\sigma } 
\left( G_{\bf k}^R-G_{\bf k}^A\right) G_{\bf k}^R\, G_{\bf k}^A\; ,
\eqno (18)
$$
and, after calculating the integral over ${\bf k}$, we find
$$     
\Delta \sigma _{xy}^{(sj)}
=-\frac{e^2\lambda _0^2}{4\pi m^*}\, {\rm Tr}\, \sum _{\bf q}\sigma _z\, P({\bf q})\; ,
\eqno (19)
$$
where $P({\bf q})$ is a diagonal matrix with the elements
$$
P_\sigma ({\bf q}) \simeq \frac{q_y^2}{D_\sigma q^2}\; .
\eqno (20)
$$
The integral over $q$ in (19) is mainly determined by the Diffuson at the 
upper limit, $q\sim (D\tau )^{-1/2}$, for which the vertex $\Gamma _x^y({\bf q})$,
Eq.~(16), was not found correctly. 

In the three-dimensional case we can estimate the integral as
$$
\int \frac{d^3q}{(2\pi )^3}
P_\sigma ({\bf q})
\simeq \frac{m^*\, k_{F\sigma }^3}{(\varepsilon _{F\sigma }\tau _\sigma )^4}\; .
\eqno (21)
$$
Combining (19),(20) with Eq.~(6), we find the relative value of the quantum
correction, 
$\Delta \sigma _{xy}^{(sj)}/\sigma _{xy}^{(sj)}\simeq (\varepsilon _F\tau )^{-4}$. 
Since the usual correction to the conductivity $\sigma _{xx}$ has the
relative magnitude of $(\varepsilon _F\tau )^{-2}$, the localization correction 
to the off-diagonal conductivity $\sigma _{xy}^{(sj)}$ turns out to be very small.

In the case of effective two-dimensionality of quantum corrections 
(when the thickness of 
magnetic film $d$ obeys inequalities $d\ll L_\varphi , L_{so}$,
where $L_\varphi $ is the phase-breaking length and $L_{so}$ is the SO scattering
length), using (19), we get 
$\Delta \sigma _{xy}^{(sj)}/\sigma _{xy}^{(sj)}\sim (\varepsilon _F\tau )^{-3}$, whereas 
$\Delta \sigma _{xx}/\sigma _{xx}\sim (\varepsilon _F\tau )^{-1}$. Thus, the
correction to $\sigma _{xy}^{(sj)}$ can be neglected for any effective 
dimensionality.

The localization corrections for the skew scattering has been calculated
earlier.\cite{langenfeld}
It should be noted, however, that the result of Ref.~[\onlinecite{langenfeld}] was 
obtained in a different model - without spin polarization of electron gas
due to the Stoner-like itinerant field $M$ (the second term in the Hamiltonian (1)) 
but with a partial polarization of spin-orbit scatterers.  To avoid possible
differences related with the choice of model, we have calculated the localization 
corrections to the AH effect due to the skew scattering from the Hamiltonian 
of Eq.~(1). 

\begin{figure}
\hskip1.5cm
\epsfig{file=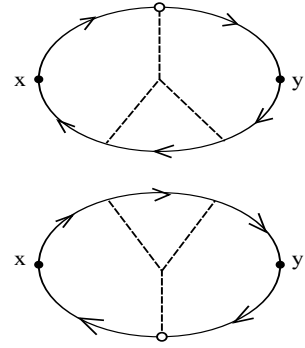,height=5.5cm,width=5cm}
\caption{Diagrams for $\sigma _{xy}^{(ss)}$ (skew scattering mechanism). 
The impurity scattering is in the third order, and the spin-orbit scattering 
amplitude is denoted by unfilled circle.}
\end{figure}

In frame of the skew scattering, we take into account the diagrams with 
the third-order corrections due to scattering from impurities, keeping the 
first order of SO-depending matrix elements. 
Without quantum corrections, the relevant diagrams for the skew scattering 
mechanism are presented in Fig.~6.\cite{crepieux}
Calculating these diagrams, we find
$$
\sigma _{xy}^{(ss)}=\frac{\pi e^2\lambda _0^2}{18\hbar }\,  
\left( k_{F\downarrow }^2\, \nu_\downarrow \, v_{F\downarrow }^2\, \tau _\downarrow 
\, \frac{\nu _\downarrow \gamma _3}{\gamma _2}\, 
-k_{F\uparrow }^2\, \nu_\uparrow \, v_{F\uparrow }^2\, \tau _\uparrow 
\, \frac{\nu _\uparrow \gamma _3}{\gamma _2}
\right) .
\eqno (22)
$$
In this formula, the dimensionless factor $\nu \gamma _3/\gamma _2$ contains
the information about both the strength of the random potential and its statistical
properties. To make it more physically sound, we introduce the average values
of the second and third powers of the mean potential at an elementary cell, 
$\langle V^2\rangle $ and $\langle V^3\rangle $. 

Taking into account that 
$\langle V^2\rangle =\gamma _2/a_0^3$ and $\langle V^3\rangle =\gamma _3/a_0^6$
(where $a_0$ is the lattice parameter), we obtain
\wide{m}{
$$
\sigma _{xy}^{(ss)}=\frac{\pi }{6}\, \frac{\langle V^3\rangle }{\langle V^2\rangle ^{3/2}}
\left[ \sigma _{xx, \downarrow }\left( \lambda _0\, k_{F\downarrow }\right) ^2
\left( \nu _\downarrow \, a_0^3\, \langle V^2\rangle ^{1/2}\right) 
-\sigma _{xx, \uparrow }\left( \lambda _0\, k_{F\uparrow }\right) ^2
\left( \nu _\uparrow \, a_0^3\, \langle V^2\rangle ^{1/2}\right) \right] .
\eqno (23)
$$
Here the dimensionless factor $\langle V^3 \rangle /\langle V^2\rangle ^{3/2}$ depends 
only on statistical properties of the random field $V({\bf r})$, whereas the
dimensionless  combination $\nu a_0^3\langle V^2\rangle ^{1/2}$ characterizes
the relative strength of the potential. 

Now we consider the diagrams with one Cooperon and three-leg impurity
vertices. There are twelve non-vanishing diagrams of the type like presented in
Fig.~7. Here the SO-dependent vertex (indicated by the white circle) 
lies on one of four possible Green
function lines. The other nine diagrams are similar to those of Fig.~7, but
differ by the location of the SO vertex. The corrections from diagrams with
Diffusons vanish.

After calculating all the diagrams, we find for the skew scattering
$$
\Delta \sigma _{xy}^{(ss)}=\frac{\pi e^2\gamma _3\lambda _0^2}{9\gamma _2}\left(
k_{F\uparrow }^2\, \nu _\uparrow ^2\, v_{F\uparrow }^2\, \tau _\uparrow ^3
\sum _{\bf q}C_\uparrow (0,{\bf q})
-k_{F\downarrow }^2\, \nu _\downarrow ^2\, v_{F\downarrow }^2\, \tau _\downarrow ^3
\sum _{\bf q}C_\downarrow (0,{\bf q})\right) .
\eqno (24)
$$
Using Eq.~(9) and calculating the integral over ${\bf q}$, the skew-scattering 
correction can be presented in the three-dimensional case as
$$
\Delta \sigma _{xy}^{(ss)}=
\frac{e^2\, \lambda _0^2}{8\sqrt{3}\, \pi \, \hbar }
\left\{
k_{F\uparrow }^2\, 
\frac{\nu _\uparrow \, \gamma _3}{\gamma _2}
\frac1{v_{F\uparrow }\, \tau _\uparrow ^{1/2}} 
\left[ \frac1{\tau _{0\uparrow }^{1/2}}
-\left( \frac1{\tau _{so\uparrow }}+\frac1{\tau _{\varphi \uparrow }}\right) ^{1/2}
\right]
-k_{F\downarrow }^2\, 
\frac{\nu _\downarrow \, \gamma _3}{\gamma _2}
\frac1{v_{F\downarrow }\, \tau _\downarrow ^{1/2}} 
\left[ \frac1{\tau _{0\downarrow }^{1/2}}
-\left( \frac1{\tau _{so\downarrow }}+\frac1{\tau _{\varphi \downarrow }}\right) ^{1/2}
\right]
\right\},
\eqno (25)
$$
where $\tau _{0\sigma }$ are some constants ($\tau _{0\sigma }\simeq \tau _\sigma $),
which can not be calculated exactly in the diffusion approximation of Eq.~(9). 

In the effectively two-dimensional case, similar calculations give us
$$
\Delta \sigma _{xy}^{(ss)}
=-\frac{e^2\, \lambda _0^2}{36\, \pi \, \hbar }
\left\{
k_{F\uparrow }^2\, 
\frac{\nu _\uparrow \gamma _3}{\gamma _2}\,  
\ln \left[\tau _\uparrow \left( \frac1{\tau _{so\uparrow }}
+\frac1{\tau _{\varphi \uparrow }}\right) \right]
-k_{F\downarrow }^2\, 
\frac{\nu _\downarrow \gamma _3}{\gamma _2}\,  
\ln \left[\tau _\downarrow \left( \frac1{\tau _{so\downarrow }}
+\frac1{\tau _{\varphi \downarrow }}\right) \right]
\right\} .
\eqno (26)
$$ 
}
Thus, the localization correction to the AH conductivity due to the skew scattering
is nonzero, in agreement with Ref.~[\onlinecite{langenfeld}].

The anomalous Hall resistivity, determined as
$$
R_{AH}=\frac{\sigma _{xy}}{\sigma _{xx}^2}\; ,
\eqno (27)
$$
acquires the corrections from both diagonal and off-diagonal conductivities
$$
\frac{\Delta R_{AH}}{R_{AH}^0}=\frac{\Delta \sigma_{xy}}{\sigma _{xy}^0}
-2\, \frac{\Delta \sigma_{xx}}{\sigma _{xx}^0}\; .
\eqno (28)
$$

Since the correction to AH conductivity in frame of the side-jump mechanism 
is zero, the total localization correction $\Delta \sigma _{xy}$ is given by (25) or (26).
The relative magnitude of this correction depends on the prevailing mechanism
of AH effect.
Using Eqs.~(7) and (22), we can find that the relative order of the AH conductivity due 
to the skew scattering or side-jump is

$$
\frac{\sigma _{xy}^{(ss)}}{\sigma _{xy}^{(sj)}} 
\simeq \frac{\nu \, \gamma _3}{\gamma _2}\left( \varepsilon _F\tau \right) .
\eqno (29)
$$

The weak-localization approach is valid as long as $(\varepsilon _F\tau )\gg 1$.
Thus, for $\nu \gamma _3/\gamma _2>1$, the skew scattering mechanism is more 
important, and the localization correction is determined by Eqs.~(25) or (26). 

\begin{figure}
\epsfig{file=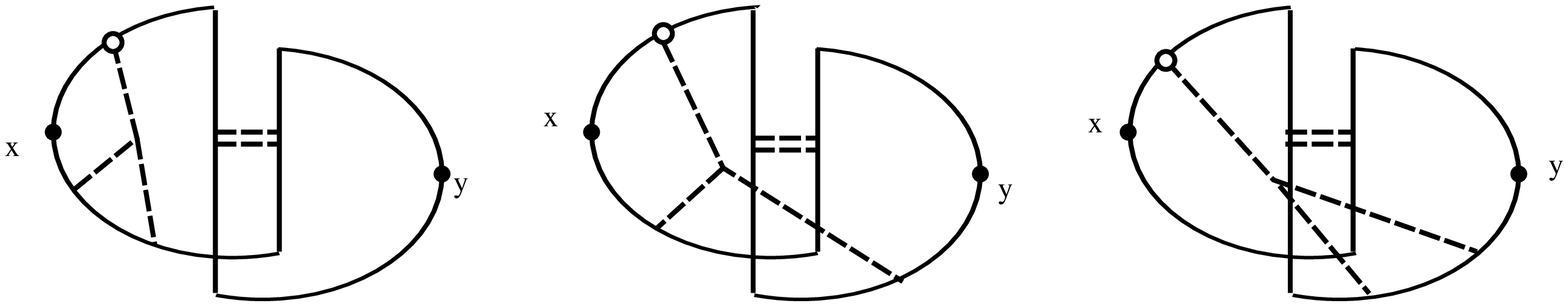,height=3cm,width=8.4cm}
\caption{First three diagrams describing localization corrections to 
$\sigma _{xy}^{(ss)}$. Other diagrams differ by locations of the spin-orbit
vertex (unfilled circle). }
\end{figure}

In the case of $(\nu \gamma _3/\gamma _2)(\varepsilon _F\tau )\ll 1$, 
the prevailing mechanism is side-jump.
Since the side-jump correction is zero, the total localization correction, 
determined by Eqs.~(25) or (26), turns out to be negligibly small:  
$\left( \Delta \sigma _{xy}^{(ss)}/\sigma _{xy}^{(sj)}\right) \simeq 
\left( \Delta \sigma _{xy}^{(ss)}/\sigma _{xy}^{(ss)}\right) 
\left[ (\nu \gamma _3/\gamma _2)\, (\varepsilon _F\tau )\right]^{1/2}
\ll \left( \Delta \sigma _{xy}^{(ss)}/\sigma _{xy}^{(ss)}\right) $. 

Collecting all together, we can formulate our final results as follows: 

(i) for the low-resistivity metals with prevailing skew scattering, the 
localization correction to AH resistivity (28) contains both parts with
$\Delta \sigma _{xy}$ (described by (25) or (26)) and $\Delta \sigma _{xx}$.
No cancellation between them is possible due to the separation of contributions 
from the different spin channels;

(ii) for the high-resistivity metals or doped semiconductors with prevailing
side-jump mechanism, the correction to $\Delta \sigma _{xy}$ is negligibly
small, so that the localization correction to AH resistivity, Eq.~(28), is exactly twice 
the relative correction to the diagonal conductivity (with the opposite sign).

These results, concerning the AH effect, differ significantly from what is 
known for the usual Hall effect, described by a Hall constant $R_H$. 
It has been shown\cite{fukuyama,alt80a,alt80b} that the localization correction 
to $R_H$, determined by an analogous formula (28), is identically zero
due to the mutual cancellation of contributions from the diagonal, 
$\Delta \sigma _{xx}^{(loc)}$, and  off-diagonal, 
$\Delta \sigma _{xy}^{(loc)}$, conductivities. 
On the other hand, considering the interaction corrections to $R_H$, it has been 
found that $\Delta \sigma _{xy}^{(int)}=0$. Thus, the total quantum corrections to 
the Hall constant are reduced to 
$\Delta R_H/R_H^0=-2\, (\Delta \sigma _{xx}^{(int)}/\sigma _{xx})$. 
 
The experiments on amorphous ferromagnetic Fe films\cite{bergmann91} 
have shown that the quantum correction to the AH resistivity (28) is double
the correction to the diagonal conductivity. 
This is in accordance with our result for the localization corrections under  
condition that the side-jump mechanism prevails. 
The latter is in agreement with the comparatively high resistivity of 
amorphous Fe films studied in Ref.~[\onlinecite{bergmann91}].

But our main argument in favor of the prevailing side-jump 
mechanism\cite{berger} is that the random field experienced by the electrons 
in amorphous films is more naturally described by a distribution $P\{V({\bf r})\}$
with nearly equal probabilities of positive and negative
deviations of the random potential $V({\bf r})$ from zero.
In such a case the parameter $\nu \gamma _3/\gamma _2$ in Eq.~(29) is small 
thanks to $\langle V^3\rangle /\langle V^2 \rangle ^{3/2}\ll 1$.

The authors of the cited works\cite{langenfeld,bergmann91} 
have given another explanation of the
measurements: suppression of localization correction to the off-diagonal
conductivity due to very strong spin-orbit scattering ($\tau _{so}\simeq \tau $),
upon the prevailing skew scattering mechanism.
Besides, the quantum corrections to the AH conductivity due to electron-electron
interaction have been also calculated for
the skew scattering in Ref.~[\onlinecite{langenfeld}], and the cancellation
of interaction corrections in $\Delta \sigma _{xy}^{(ss,\, int)}$
has been proved. It should be noted, however, that
the Hartree diagrams were not taken into account in this calculation. 

In conclusion, we have shown that the role of localization corrections is quite
different for the skew scattering and side-jump mechanisms of AH effect.
We suggest that the experimental results of Ref.~[\onlinecite{bergmann91}] can be 
interpreted as a relative smallness of the localization correction to the off-diagonal 
conductivity upon the prevailing side-jump mechanism.  

V.D. is thankful to J. Barna\'s for numerous discussions and critical reading of
the manuscript and to Polish KBN for a partial support under Grant No.  5 P03B 091 20.

\end{multicols}

\end{document}